\RequirePackage{fix-cm}
\documentclass[twocolumn]{article}       
\usepackage{graphicx}
\usepackage{color}
\usepackage{amsmath,amssymb}
\usepackage{mathtools}
    \mathtoolsset{showonlyrefs}

\DeclareMathAlphabet{\mathbf}{OT1}{cmr}{bx}{it}
\DeclareMathAlphabet{\bfit}{OT1}{cmr}{bx}{it}

\newcommand {\0} {\textbf{0}}    
\newcommand {\Ec}  {\mathcal{E}}
\newcommand {\Ab} {\mathbf{A}}
\newcommand {\pb} {\mathbf{p}}
\newcommand {\tb} {\mathbf{t}}
\newcommand {\Mb} {\mathbf{M}}

\newcommand{\Mathematica}{\textsl{Mathematica}\textsuperscript{\resizebox{!}{0.8ex}{\textregistered}}\ }
\newcommand{\Matlab}{\textsl{Matlab}\textsuperscript{\resizebox{!}{0.8ex}{\textregistered}}\ }

\begin{document}
\title{\vskip-3cm\bf On the compact wave dynamics of tensegrity beams in multiple dimensions}



\author{
Andrea Micheletti\\[10pt]
DICII\\ University of Rome Tor Vergata\\ Italy\\ {\em micheletti@ing.uniroma2.it}      
\and
Giuseppe Ruscica\\[10pt] DISA\\ University of Bergamo\\ Italy\\ {\em giuseppe.ruscica@unibg.it}         
\and
Fernando Fraternali\\[10pt] DICIV\\ University of Salerno\\ Italy\\ {\em f.fraternali@unisa.it} 
}




%

\date{\ }


\maketitle

\begin{abstract}
This work presents a numerical investigation on the nonlinear wave dynamics of tensegrity beams in 1D, 2D and 3D arrangements. The simulation of impact loading on a chain of tensegrity prisms and lumped masses
allows us to apply on a smaller scale recent results on the propagation of compression solitary waves in 1D tensegrity metamaterials. 
Novel results on the wave dynamics of 2D and 3D beams reveal~-- for the first time ~-- the presence of compact compression waves in two- and three-dimensional tensegrity lattices with slender aspect ratio. The dynamics of such systems is characterized by the thermalization of the lattice nearby the impacted regions of the boundary. The portion of the absorbed energy moving along the longitudinal direction is transported by compression waves with compact support. Such waves emerge with nearly constant speed, and slight modifications of their spatial shape and amplitude, after collisions with compression waves traveling in opposite direction. The analyzed behaviors suggest the use of multidimensional tensegrity lattices for the design and  additive manufacturing of novel sound focusing devices.

\end{abstract}

\vskip5pt



\section{Introduction}
\label{intro}
Lattice metamaterials are periodic systems that tassellate spatial domains with structured building blocks, in order to form engineered materials featuring exceptional values of key properties. The latter include negative effective elastic moduli and mass density, frequency bandgaps, auxetic response, exceptional stiffness/weight and strength/weight ratios, to name but a few examples (refer, e.g., to \cite{Liu2000}--\cite{Phani2017}, and references therein).
Originally, the unconventional behaviors of mechanical metamaterials were achieved through linear response of the unit cells, and an ad-hoc design of the internal architecture of the system, obtaining an overall mechanical response that goes beyond that of the constituent materials.
Nonlinear metamaterials are nowadays emerging as mechanical systems with highly tunable response, which is induced by nonlinearities linked to large displacements/strains, soft-modes and/or mechanical instabilities \cite{Ruzzene2010,Ruzzene2014,Bertoldi2017}. 
Different studies available in the literature have shown that elastically hardening (or stiffening) discrete systems support compressive solitary waves \cite{frieseckeMatthies2002,Theocharis2013,Nesterenko_2001}, while elastically softening systems support the propagation of rarefaction solitary waves under initially compressive impact loading \cite{Nesterenko_2001,Herbold2012}. 

Solitary waves are mechanical waves whose waveform shows one global peak propagating with constant size and shape, which progressively decays moving away from the peak (refer, e.g., to \cite{Nesterenko_2001,Rosenau1987} and references therein). Solitons are special solitary waves that  emerge unmodified after collisions with other solitons, with exception to a phase shift \cite{Rosenau2005}. Particularly interesting is the case of solitons with finite span (or wavelenegth)--usually refereed to as `compactons' \cite{Rosenau2005}. Solitary waves characterized by different, symmetric, antisymmetric, and cusped profiles  are actively investigated in the literature \cite{ndyn5,ndyn6,ndyn8}, which is also populated by multifaceted studies on the stability of such waves \cite{ndyn6,ndyn1,ndyn2}. From the engineering point of view, solitary wave dynamics has been proven to be useful for the construction of a variety of novel acoustic devices, including: acoustic band gap materials; shock protector devices; acoustic lenses; and energy trapping containers (refer, e.g., to \cite{Theocharis2013} and references therein). 
Particularly interesting is the solitary wave dynamics of discrete systems alternating  tensegrity units with lumped masses (1D `tensegrity metamaterials') \cite{25,28,Davini_2016,IJSS2018}. Tensegrity systems are known as prestressable truss structures whose stiffness matrix is composed of material and geometric terms \cite{Skelton_2010}. The material term stems from the material, geometric and size properties of the members, while the geometric part of the stiffness matrix depends on the state of stress acting in the current configuration of the structure, and the change of geometry of the system when moving from such a configuration \cite{Skelton_2010,Micheletti2013,SMS15,Amendola_2015,27,1}. In correspondence with infinitesimal mechanisms, which are customary in tensegrity structures, the stiffness of the system is dominated by the geometric term, and it may happen that a given mechanism is stabilized by the internal self-stress, producing a stiffening-type response (prestress-stable structures) \cite{Micheletti2013,SMS15}. The influence of the geometrical nonlinearities on the solitary wave dynamics of one-dimensional tensegrity metamaterials has been diffusely studied in \cite{25,28,Davini_2016}.

The present study investigates the propagation of mechanical waves with compact support in multidimensional tensegrity beams with stiffening-type elastic response, through suitable extension and generalization of the one-dimensional studies presented in \cite{25,28,Davini_2016}. We name `beams' the analyzed systems because their longitudinal dimension is markedly greater than their transverse size. These systems are obtained through assemblies of tensegrity prisms and lumped masses, such that the masses move only in the longitudinal direction (1D chains), as well as through slender arrangements of 2D and 3D unit cells. We begin by reviewing the origin and nature of the stiffening response of 1D, 2D and 3D tensegrity lattices, which arise at the unit level and at the interface between different units (Sect. \ref{stiffening}). Next we present the mechanical  model and the numerical integration procedure that are employed in this work to simulate the wave dynamics of tensegrity lattices (Sect. \ref{model}).
In Sect. \ref{sec:1Dwaves}, we review the propagation of compact compression waves in 1D tensegrity chains, by complementing the study presented in \cite{25} with new results on a small-scale system. 
Sects. \ref{sec:2Dwaves},  \ref{sec:3Dwaves} present numerical experiments on the propagation of mechanical waves in 2D and 3D tensegrity beams, which are impacted by impulsive compressive disturbances on one or two opposite edges of the boundary. The given results in two- and three-dimensions lead us to generalize previous results dealing with the one-dimensional wave dynamics of tensegrity mass-spring systems \cite{25,28,Davini_2016}, and to prove, for the first time, the presence of compact compression waves in 2D and 3D arrangements of tensegrity units.  Our impact simulations also reveal some distinctive features of 2D and 3D systems, which are related to thermalization effects near the impacted zones of the boundary, and the spatial distribution of the transported energy among the bars and cables at the wavefront.
We end the present work in Sect. \ref{conclusions} with some concluding remarks and directions for future work.


\section{Stiffening response of tensegrity lattices} \label{stiffening}

The theory of nonlinear waves in discrete 1D materials presented in \cite{Nesterenko_2001} predicts that particulate systems featuring power-law interactions with exponent $n$ greater than one (elastically hardening - or stiffening - systems) support energy transport through compression solitary waves, at the steady state. In weakly precompressed systems, the characteristic phase speed $V_s$ and the spatial length $L_n$ of such waves obey the following laws (cf. Sect. 1.10 of  \cite{Nesterenko_2001}) 

\begin{equation}
V_s \ = \  c_n \ \sqrt{\frac{2}{n+1}} \ \varepsilon_m^{\frac{n-1}{2}} \,
\label{eq:Vs}
\end{equation}

\begin{equation}
L_n \ = \  \frac{\pi a}{n-1} \ \sqrt{\frac{n(n+1)}{6}} \,
\label{eq:Ln}
\end{equation}

\noindent where $a$ is the lattice constant (i.e., the particle spacing when the system is at rest), $\varepsilon_m$ is the maximum axial strain experienced by the system under the applied external excitation, and $c_n$ is a constant with dimensions of a speed. Such a  quantity is related to the long-wave sound speed of the system $c_0$ through

\begin{equation}
c_0 \ = \  c_n \ \sqrt{n} \ \varepsilon_0^{\frac{n-1}{2}} \,
\label{eq:c0}
\end{equation}

\noindent $\varepsilon_0$ denoting the axial strain produced by an external precompression force $F_0$. Eqn. \eqref{eq:c0} shows that the speed of sound $c_0$ approaches zero (i.e., the material behaves as a `sonic vacuum' \cite{Nesterenko_2001}) when it results $\varepsilon_0=0$, implying that the system is not precompressed at the initial state.  For what concerns the width $L_n$ of the solitary waves, Eqn. \eqref{eq:Ln} shows that that the traveling solitons span 5 particles ($L_n = 4.97 a$) for $n=1.5$ (e.g., in a granular medium with Hertzian interaction forces), and about 3 particles ($L_n = 2.22 a$) for $n=3.0$. Different is the case of discrete power-law systems featuring elastically softening response (power-law materials with $n<1$), which instead support the propagation of rarefaction solitary waves \cite{Nesterenko_2001}).
Alternative approaches to the dynamics of mechanical system exhibiting power-law nonlinearities are presented  in  \cite{Rosenau1987,ndyn7,ndyn4}.

The following sections illustrate the stiffening-type mechanical responses of various examples of tensegrity systems, which originate at the unit level (Sect.  \ref{unit_stiffening}), or at the interface between different units (Sect.  \ref{interface_stiffening}). The analyzed behaviors are responsible for the peculiar dynamics of the systems analyzed in the subsequent Sects. \ref{sec:1Dwaves}-\ref{sec:3Dwaves}.

\subsection{Unit-level stiffening response} \label{unit_stiffening}

\begin{figure}[htbp]
\begin{center}
\includegraphics[width=0.9\linewidth]{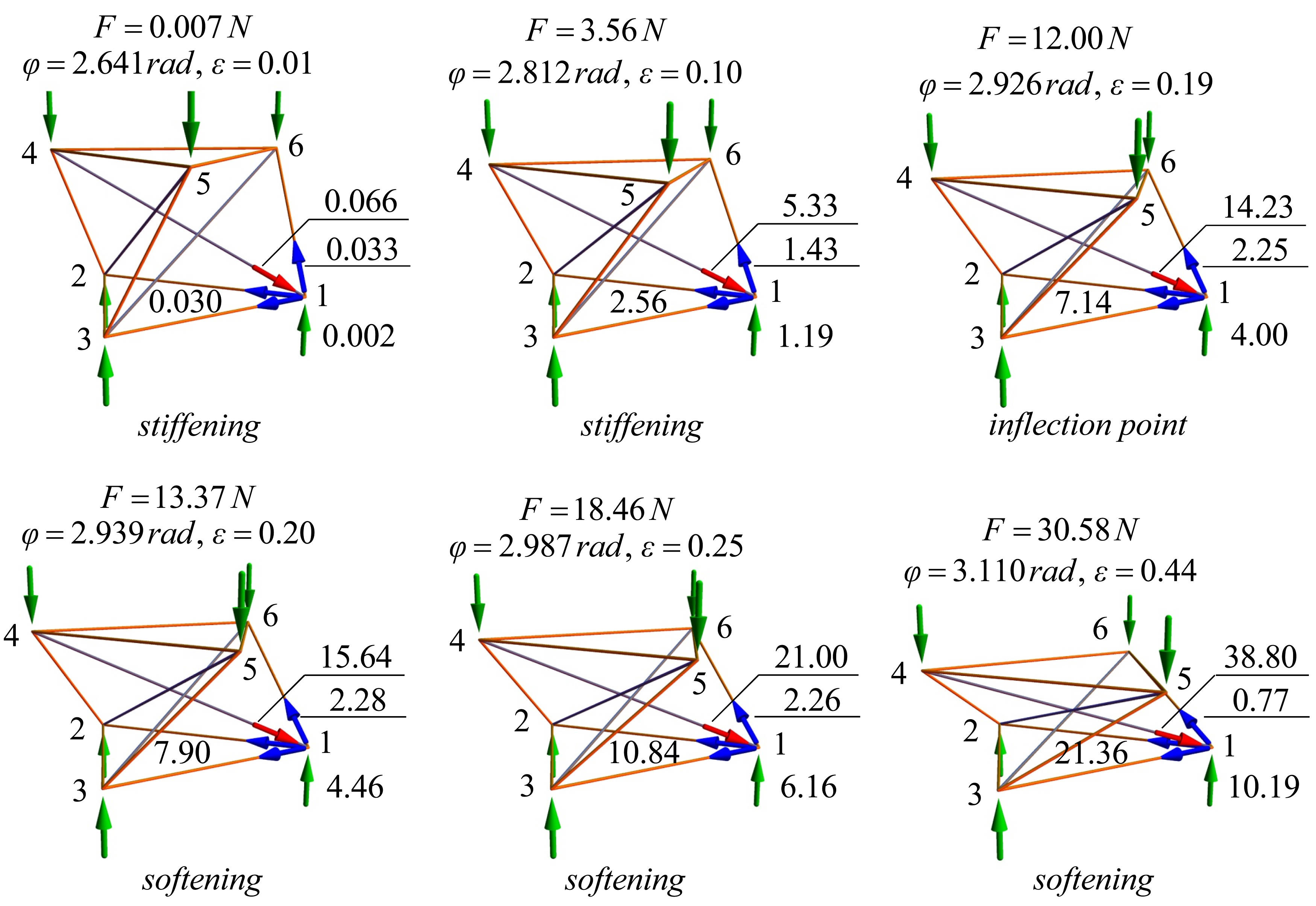}
    \caption{Sequence of deformed configurations of a t-prism under uniform compression loading.}\label{Prism_deformation}
\end{center}
\end{figure}

\begin{figure}[htbp]
\begin{center}
\includegraphics[width=0.925\linewidth]{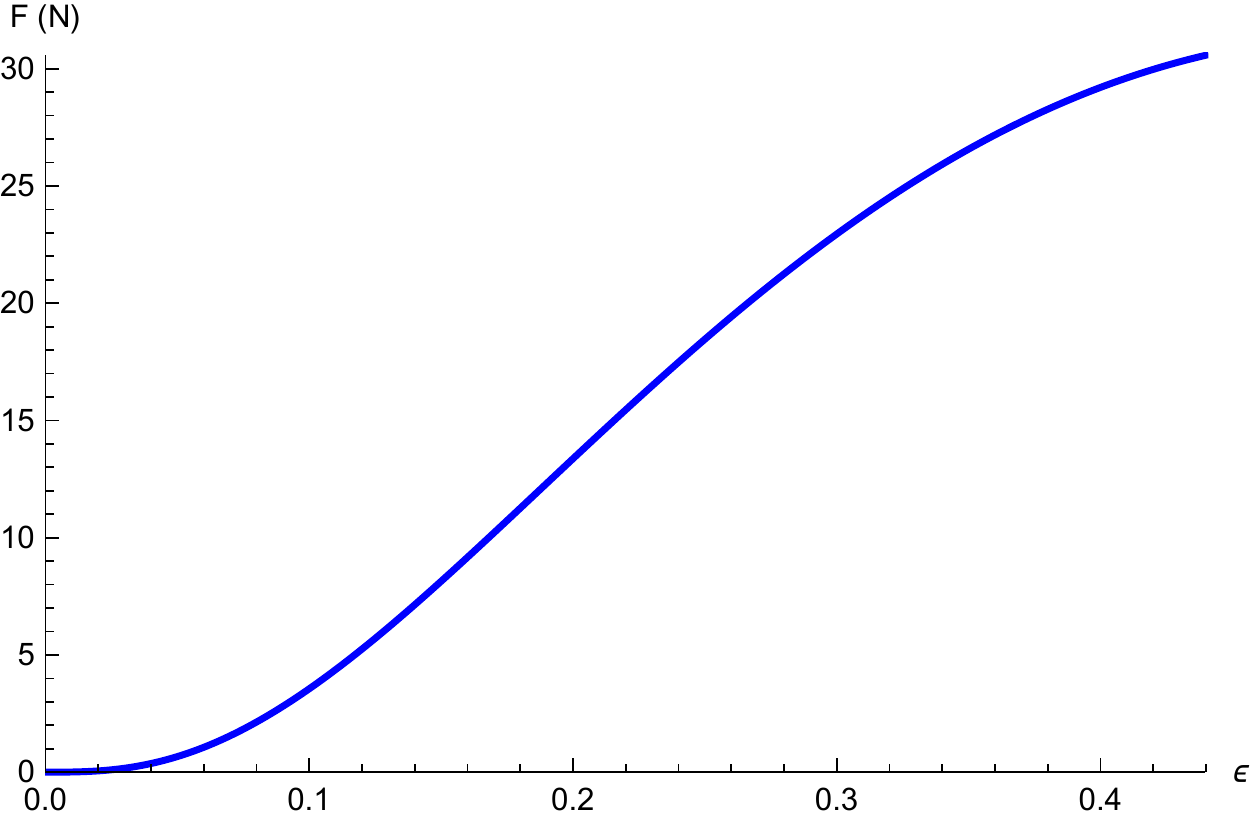}
    \caption{Axial force vs. axial strain response of a t-prism under uniform compression loading.}\label{Fepsplot}
\end{center}
\end{figure}

A well-studied, unit-level stiffening response is observed under compression loading in minimal regular tensegrity prisms \cite{Skelton_2010}, hereafter simply referred to as t-prisms. Let us refer to the small-scale model analyzed in \cite{28,IJSS2018} that is formed by 0.8 mm circular bars made of the titanium alloy Ti$_6$Al$_4$V ($120$ GPa Young's modulus; $4.42$ g/cm$^3$ mass density), and 0.28 mm Spectra fibers ($5.48$ GPa Young's modulus;  $0.98$ g/cm$^3$ mass density). Fig. \ref{Prism_deformation} shows a sequence of deformed configurations of such a structure from the freestanding configuration featuring height $h_0=5.41$ mm \cite{28,IJSS2018}, under zero internal self-stress. The configurations depicted in Fig. \ref{Prism_deformation}  have been determined through the path-following approach presented in \cite{27}. The axial force $F$ vs. axial strain $\varepsilon = (h_0 - h) / h_0$ response shown in Fig. \ref{Fepsplot} highlight an initially stiffening response of the t-prism (i.e., tangent stiffness growing from zero with the axial strain), followed by a softening response (concavity facing downward) under large axial strains.

It is useful to fit the $F$ vs. $\varepsilon$ response in Fig. \ref{Fepsplot} with a power-law of the following type

\begin{equation}
F \ = \  c \ \varepsilon^n \ .
\label{eq:powerfit}
\end{equation}

\begin{figure}[htbp]
\begin{center}
\includegraphics[width=0.95\linewidth]{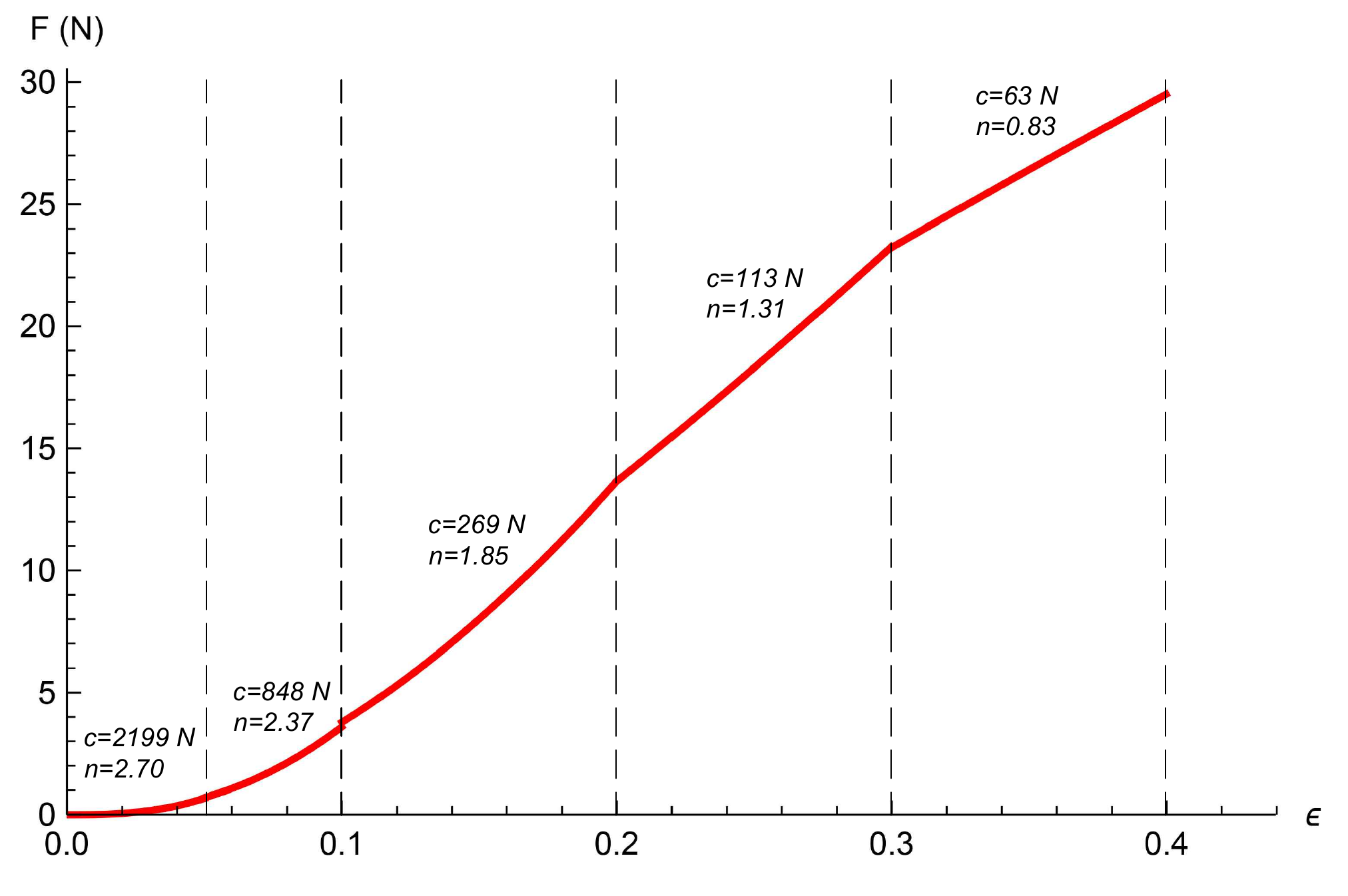}
    \caption{Power-law fitting of the axial force vs. axial strain response of a t-prism.}\label{Fepsfit}
\end{center}
\end{figure}

\noindent The use of the  {\tt{FindFit}} function of \Mathematica leads us to the results graphically illustrated in Fig. \ref{Fepsfit}, which predict: $c=2199 \ \mbox{N}, n=2.70$ for $\varepsilon \in [0, 0.05]$; $c=848 \ \mbox{N}, n=2.37$ for $\varepsilon \in [0.05, 0.10]$; $c=269 \ \mbox{N}, n=1.85$ for $\varepsilon \in [0.10, 0.20]$; $c=113 \ \mbox{N}, n=1.31$ for $\varepsilon \in [0.20, 0.30]$; and $c=63 \ \mbox{N}, n=0.83$ for $\varepsilon \in [0.30, 0.40]$. The above results reveal that both $c$ and $n$ decrease with the applied strain, and that one obtains a power-law with exponent lower than one for $\varepsilon > 0.30$. Differently, when $\varepsilon$ is lower than 30\%, one instead observes power-law fits of the $F-\varepsilon$ response with exponents greater that one. Such a stiffening behavior induces a dynamic response characterized by the formation and propagation of solitary waves with spatial width varying from 2.4 $h_0$ ($\varepsilon \in [0, 0.05]$) to 7.08 $h_0$ ($\varepsilon \in [0.20, 0.30]$), according to the theory presented in \cite{Nesterenko_2001} (cf. Eqn. \eqref{eq:Ln}).

\subsection{Interface-level stiffening response} \label{interface_stiffening}

\begin{figure}[htbp]
\begin{center}
\includegraphics[width=\linewidth]{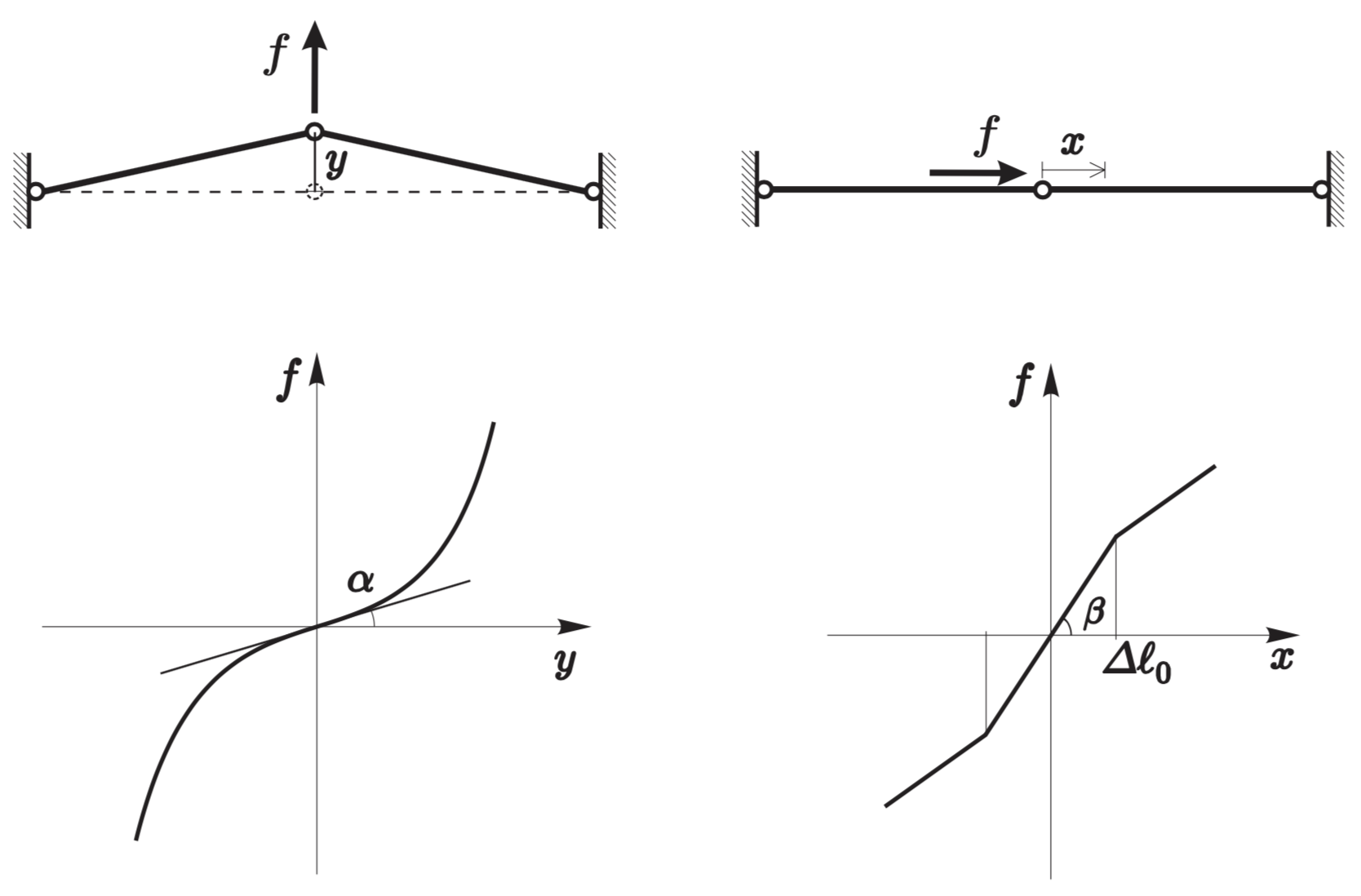}
    \caption{Nonlinear response of a two-string system.}\label{twoedges}
\end{center}
\end{figure}

The stiffening behaviors that originate from interpenetration and interlocking mechanisms between adjacent unit cells  can be analyzed by studying the elementary two-string system depicted in Fig.~\ref{twoedges}. Starting with the aligned configuration, where the two horizontal strings are prestressed with a certain tensile force, one can distinguish two basic behaviors \cite{Micheletti2013}. The response to a vertical load $f$ (acting along the internal mechanism) is well approximated by a cubic law with inflection point at the origin, as we shall see in short (Fig.~\ref{twoedges}, bottom-left). The slope ${\rm tan}\,\alpha$ of the tangent at the origin is proportional to the prestressing force. Under a horizontal load (orthogonal to the mechanism, Fig.~\ref{twoedges}, top-left), the response of the analyzed system is instead linear. Assuming no-compression response of the cables (i.e., zero buckling load), this linear response will be characterized by a marked change of slope when the applied horizontal load exceeds the given pretension, leading one of the two cables to go slack (Fig.~\ref{twoedges}, bottom-right). The response to an inclined force is a (nonlinear) combination of these two basic behaviors. 

Let us now focus our attention on the system depicted in Fig.~\ref{twoedges}-left, assuming that it is composed of two identical linear-elastic strings with rest length $\bar l$, and initial length $l_0$. Denoting the vertical displacement of the center node by $y$, we compute the deformed length of each string as follows: $l(y)=\sqrt{l_0^2+y^2}$. The total potential energy of the system ($\Ec_{\rm total}=\Ec_{\rm elastic}+\Ec_{\rm loads}$) can be cast into the following form

\begin{equation}
\Ec_{\rm total}\,=\,k\,\big( l(y) - \bar l \big)^2\,-\,f\,y\,,
\label{eq:Ec}
\end{equation}

\noindent where we denoted the elastic constant of each string by $k$. The stationarity condition $\Ec'_{\rm total}=0$ allow us to compute the displacement $y$ at equilibrium, for each given value of the load $f$ (here and in what follows, primes denotes derivatives with respect to $y$) . It is worth noticing the the stability condition $\Ec''_{\rm total}>0$, always holds when it results $\bar l < l_0$. It is also easily verified that the stationarity of the potential energy \eqref{eq:Ec} leads  us to the following relation between the transverse force and the transverse displacement 

\begin{equation}
f(y)=2\,k\,\Big(l(y)- \bar \l \Big)\frac{y}{l(y)}\ .
\label{eq:ffunc}
\end{equation}

\noindent By taking the Taylor series expansion of this expression up to the fourth order, we finally obtain the following $f$ vs. $y$ response

\begin{equation}
f(y)=\frac{2\,k\,(l_0-\bar l)}{l_0}\,y\,+\,k\,\frac{\bar l}{l_0^3}\,y^3+o(y^5),
\label{eq:f-y}
\end{equation}

\noindent which shows a cubic-type profile under moderately large displacements $y$. It is easily observed that the linear term in this relation vanishes when it results $l_0=\bar l$, that is, when the prestress is zero.

\begin{figure}[htbp]
\begin{center}
\includegraphics[width=1.0\linewidth]{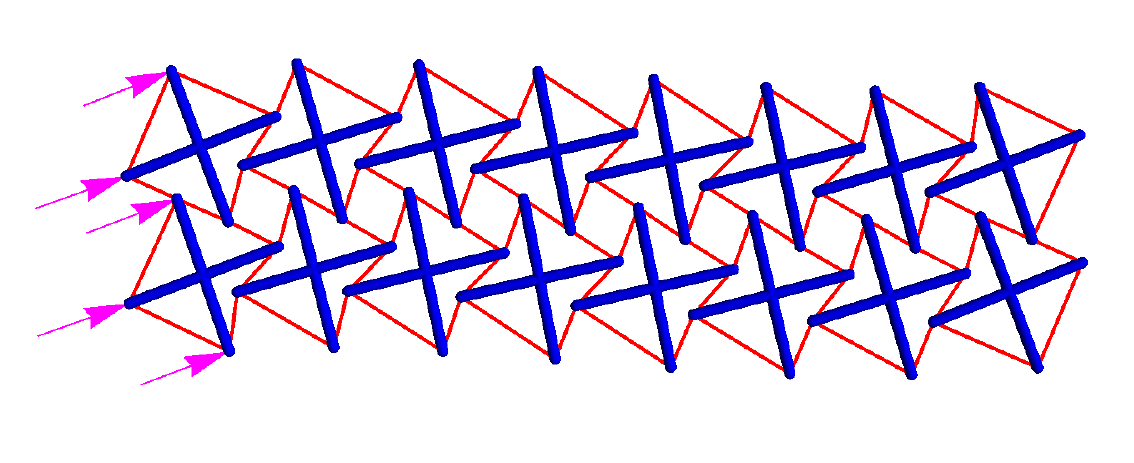} \\
\includegraphics[width=0.65\linewidth]{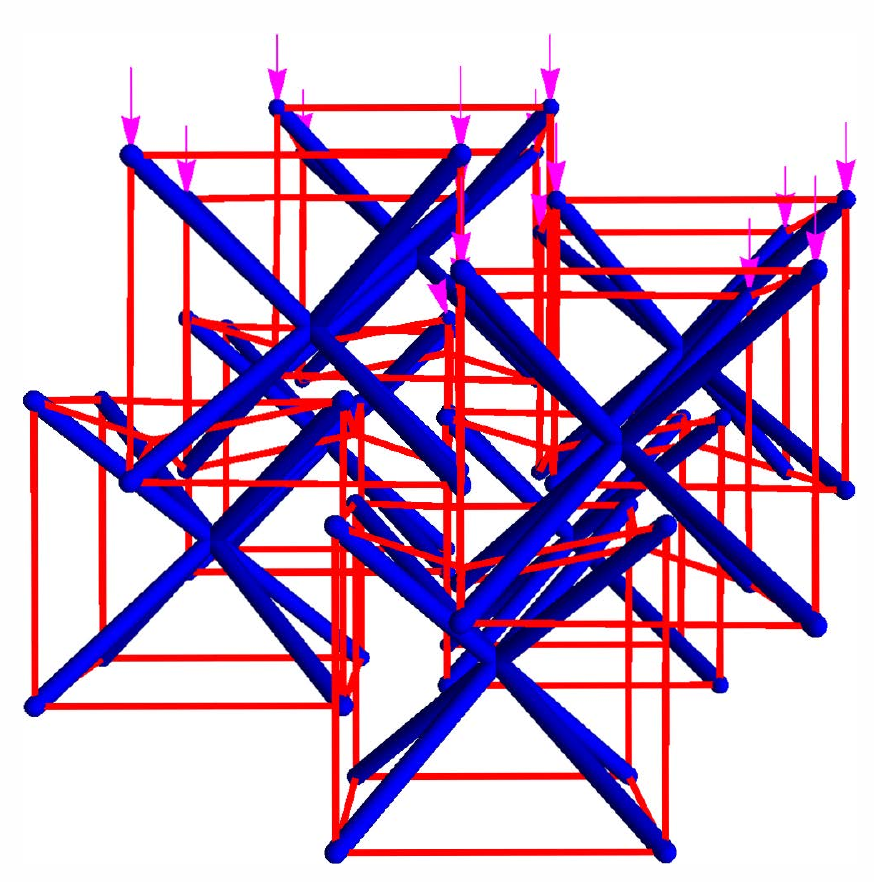}
    \caption{Two- and three-dimensional tensegrity systems undergoing interpenetration of the unit cells under compression loading. Top: a $2\times8$ assembly of square cells. Bottom: a $2\times2\times2$ assembly of cubic cells.}\label{assemblages}
\end{center}
\end{figure}

\begin{figure}[htbp]
\begin{center}
\includegraphics[width=0.8\linewidth]{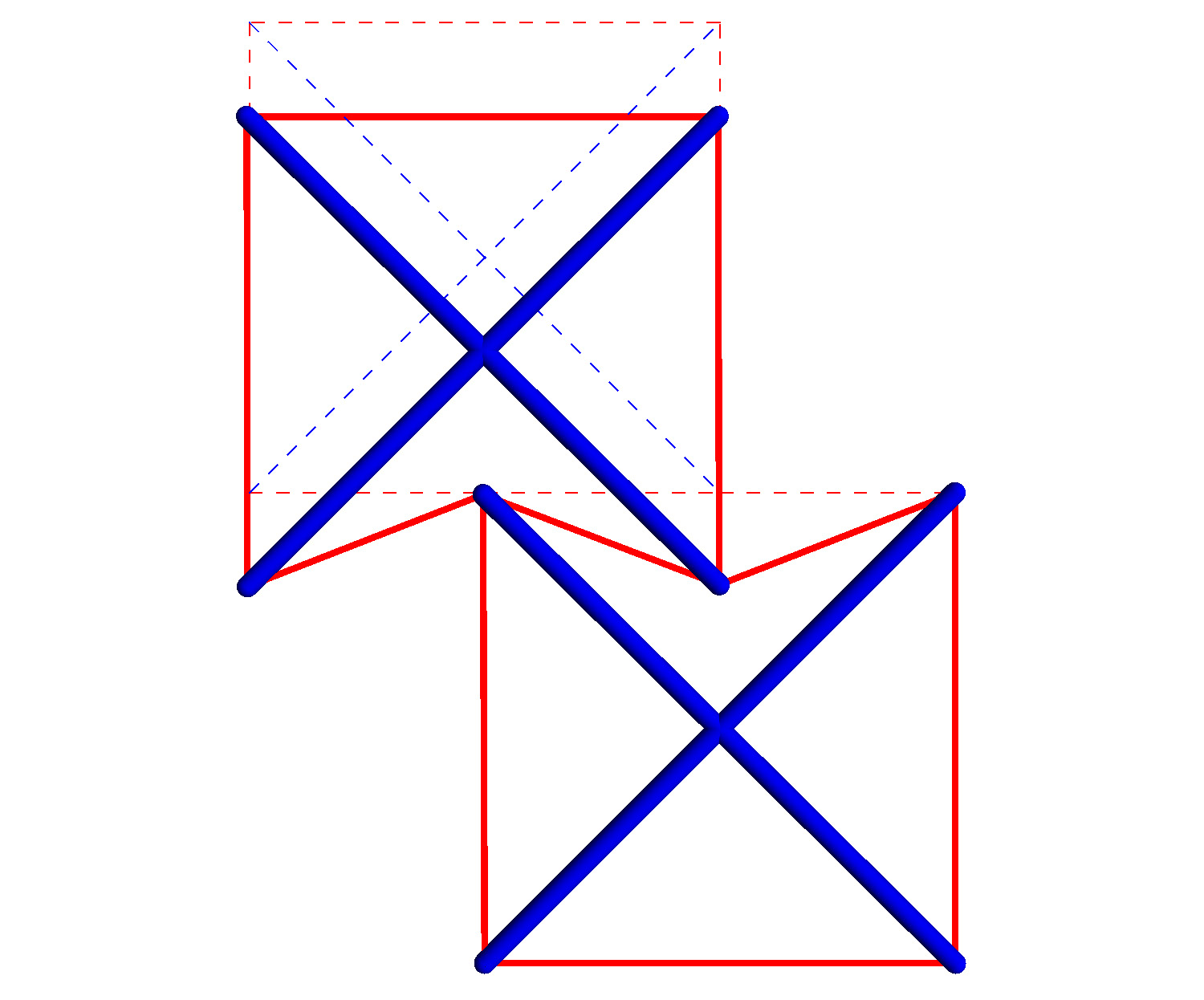}
    \caption{Interpenetration mechanism characterizing the mechanics of the 2D system in Fig. \ref{assemblages}.}\label{compression2d}
\end{center}
\end{figure}

\begin{figure}[htbp]
\begin{center}
\includegraphics[width=0.8\linewidth]{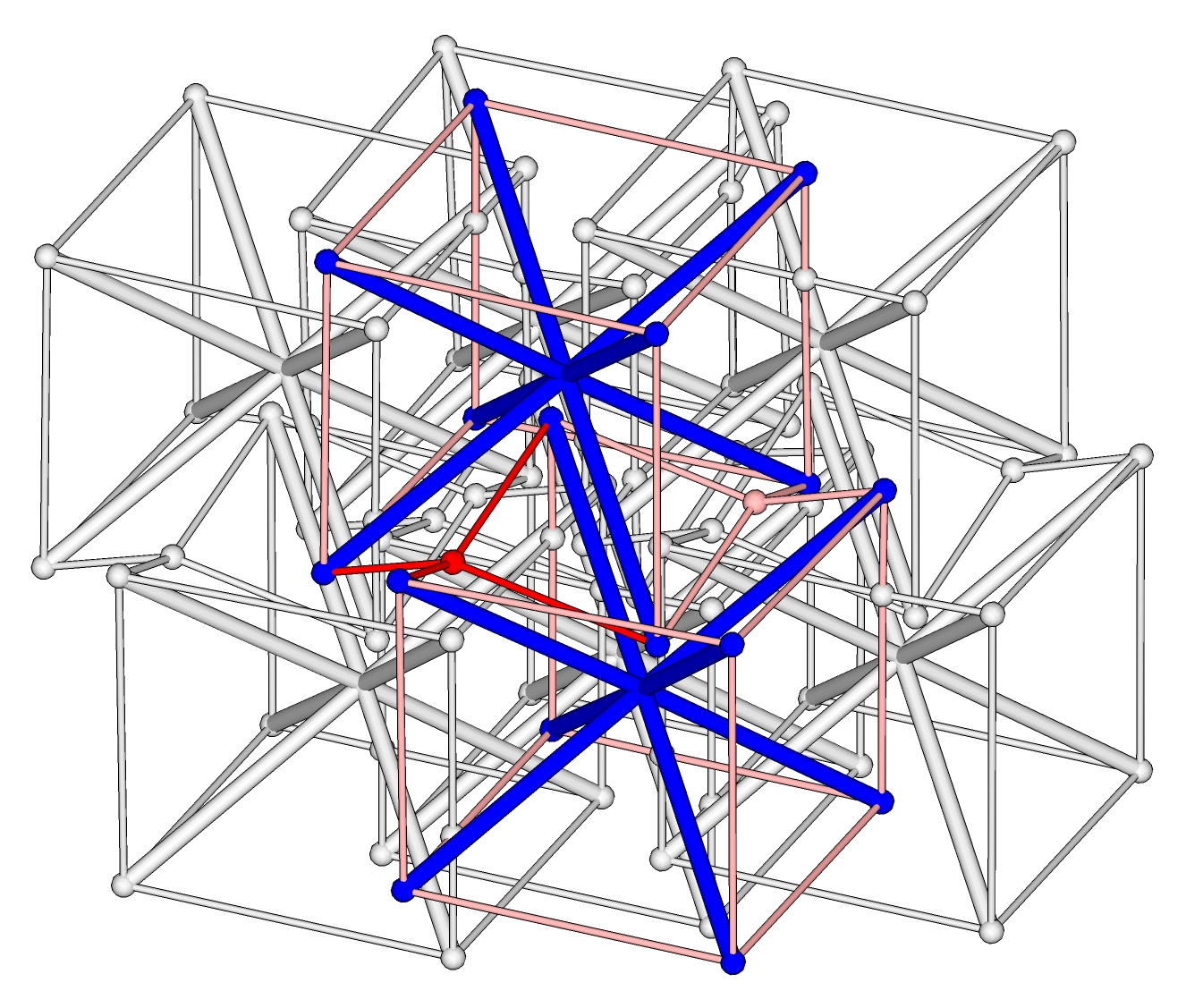}
    \caption{Entaglement mechanism characterizing the reponse of the 3D system in Fig. \ref{assemblages}.}    \label{compression3d}
\end{center}
\end{figure}

The above stiffening behavior is at the basis of the nonlinear response of 2D and 3D tensegrity systems that undergo interpenetration of adjacent units cells. Let us refer, e.g., to the systems shown  in Fig. \ref{assemblages}. The units cells forming such systems do not exhibit stiffening response as isolated structures. However, when two adjacent cells interpenetrate each other, due, e.g, to compression loading, either a bar pushes against a cable (Fig. \ref{compression2d}) or a cable gets entangled with an adjacent one (Fig. \ref{compression3d}). Such interlocking phenomena induce deformation modes of the strings that replicate the mechanism illustrated in Fig.~\ref{twoedges}-left, giving rise to a marked stiffening behavior of the overall structure. 


%


\section{Mechanical modeling of impact simulations} \label{model}

The numerical simulations presented in the following sections analyze the impact dynamics of tensegrity lattices in the large displacement regime. We model the examined lattices as collections of linear springs, with bars carrying either compressive or tensile forces, and cables carrying only tensile forces. In addition, the cables are supposed to be massless, and the mass of the system is lumped at the bars endpoints (cf. the 2D and 3D systems analyzed in Sects. \ref{sec:2Dwaves},\ref{sec:3Dwaves}), or at massive discs connecting the different units (cf. the 1D system analyzed in Sect. \ref{sec:1Dwaves}).
Under such assumptions, the equations of motion of the employed mechanical model can be written into the following matrix form ({\em cf.} \cite{Sultan2003}).
\begin{equation} \label{motioneqn}
\Ab(\pb) \tb(\pb) + \Mb \ddot \pb = \0,
\end{equation}
where $\pb$ is the vector of nodal positions, $\Mb$ is the (constant) mass matrix, $\Ab(\pb)$ is the equilibrium operator \cite{pca86}, and $\tb(\pb)$ is the vector containing the axial forces of all members. The axial force in the $i$-th member is computed as follows

\begin{equation}
t_i\,=\,\tilde k_i(\pb)\,(l_i(\pb)-\bar l_{i})\,, 
\label{ti}
\end{equation}

\noindent where $l_i(\pb)$ is the distance between the endpoints of such an element, and $\bar l_{i}$ is its rest length. 
In agreement with the adopted mechanical assumptions, and letting $k_i$ denote the stiffness constant of the $i$-th member, in Eqn. \eqref{ti} we write $\tilde k_i(\pb)=k_i$ when the $i$-th element is a bar or a cable with $l_i(\pb) \geq \bar l_{i}$, and $\tilde k_i(\pb)=0$ when instead the $i$-th element is a cable with $l_i(\pb)< \bar l_{i}$.
The initial conditions of Eqn. \eqref{motioneqn} are obtained by prescribing a certain initial velocity to the nodes subject to impact loading. The solution $\pb(t)$ is obtained by numerically integrating \eqref{motioneqn}  using an in-house developed \Matlab script. Such a code makes use of the built-in routine ode45 for the integration of systems of ordinary differential equations. 

For what concerns the adopted geometrical and mechanical properties, the 1D system analyzed in Sect. \ref{sec:1Dwaves} makes use of the t-prisms diffusely described in Sect. \ref{unit_stiffening}.
The 2D and 3D systems analyzed in Sects. \ref{sec:2Dwaves}, \ref{sec:3Dwaves} instead employ the units illustrated in Fig. \ref{assemblages}, and the geometric and mechanical properties reported in Table~\ref{tab:1}.  
It is worth noting that the unit cell edge of such systems  has length  of $20$\,mm (small-scale systems). The bars have circular cross-section with $1.75$\,mm diameter and are made of the same titanium alloy employed for the bars of the 1D system studied in Sect. \ref{sec:1Dwaves}. The cables consist of Nylon 12 fibers with $0.25$\,mm diameter,  $500$\,MPa  Young's modulus and yield strain up to over $30$\% ($1.03$\,g/cm$^3$ mass density) \cite{nylon12}.
Additional spherical masses made of lead (radius $2.5$\,mm, mass density $11.34$\,g/cm$^3$) are added at each bar's endpoint. The initial prestrain of cables, $(l_i-\bar l_i)/\bar l_i$, is prescribed approximatively equal to zero ($10^{-5}$), so that such members carry a negligible initial axial force before the system is impacted. As a consequence, the analyzed systems approximatively behave as sonic vacua, with nearly zero sound speed for infinitesimal oscillations along zero stiffness mechanisms \cite{Nesterenko_2001}.

\begin{table}
\caption{Geometrical and mechanical properties of 2D and 3D systems.}
\label{tab:1}       
\begin{tabular}{lll}
\hline\noalign{\smallskip}
quantity & value & units  \\
\noalign{\smallskip}\hline\noalign{\smallskip}
cell side & 20 & mm \\
bar diameter & 1.75 & mm \\
cable diameter & 0.25 & mm \\
Titanium Young's modulus & 120 & Gpa \\
Titanium mass density & 4.42 & g/cm$^3$ \\
Nylon 12 cables Young's modulus  & 0.5 & Gpa \\
Lead mass density & 11.34 & g/cm$^3$ \\
spherical mass radius & 2.5 & mm \\
additional nodal mass & 0.74 & g \\
\noalign{\smallskip}\hline
\end{tabular}
\end{table}

We wish to remark that the mechanical response of the structures studied in the present work is not affected by dissipative phenomena, being characterized by nonlinearities originating only from geometric effects and unilateral (no-compression) response of the cables. The analyzed systems therefore conserve their mechanical energy in correspondence of all the impact simulations presented throughout the paper.

\section{Compression solitary waves on 1D chains}
\label{sec:1Dwaves}

\begin{figure}[htbp]
\begin{center}
\includegraphics[width=\linewidth]{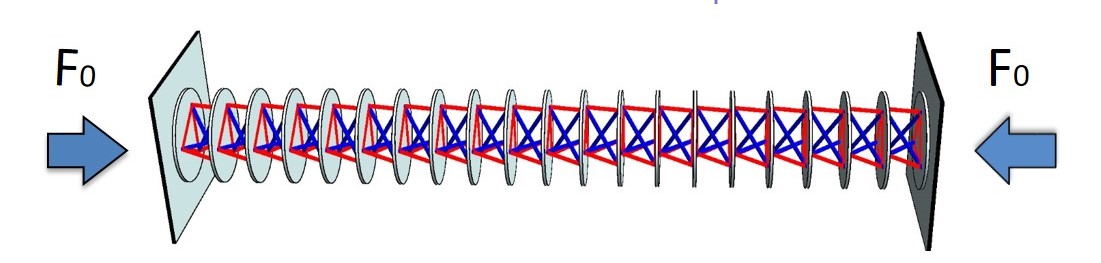}
    \caption{One-dimensional chain alternating tensegrity prisms and lumped masses.}    \label{chain1d}
\end{center}
\end{figure}

Let us begin the analysis of the nonlinear dynamics of tensegrity lattices by studying the response to impact loading of a small scale, weakly precompressed chain formed by t-prisms  alternating with lumped masses (Figure~\ref{chain1d}). The analyzed chain exhibits stiffening-type elastic response, and features the geometrical and mechanical properties reported in Sect. \ref{unit_stiffening}. A similar small-scale system has been studied in  \cite{28}  under a different, softening-type elastic regime induced  by larger precompression forces $F_0$. The lumped masses are composed of 700 lead discs with mass $m=24.82$ g each. The mass $m_0$ of a single prism is much lower than the mass of the lead discs, and amounts to $0.08$ g.

\begin{figure}[htbp]
\begin{center}
\includegraphics[width=\linewidth]{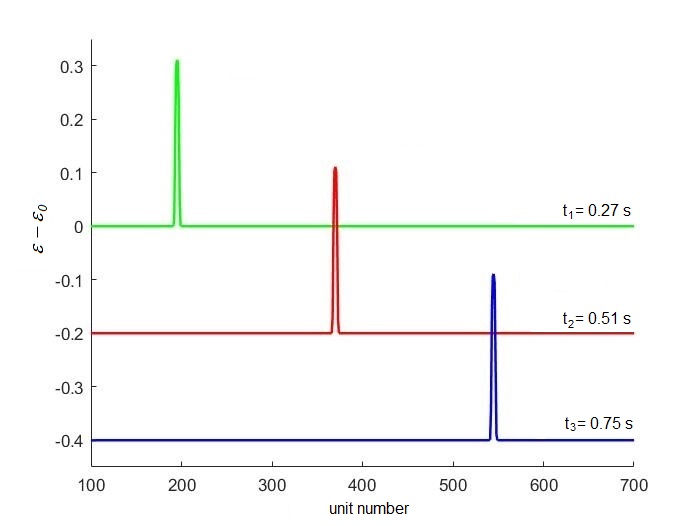}
    \caption{Propagation of compression solitary waves in a tensegrity chain (the strain is offset for visual clarity).}\label{waveplot1d}
\end{center}
\end{figure}

As in \cite{28}, 
we assume frictionless unilateral contact between the tensegrity units and the massive discs forming the chain, allowing the masses to move only in the axial direction (1D mass-spring system). The analyzed structure is subject to a small precompression force $F_0$, which induces an initial (global) strain $\varepsilon = 0.01$ (cf. Fig. \ref{Prism_deformation}), and produces a stiffening-type response of the units (cf. Sect. \ref{unit_stiffening} and Figs. \ref{Fepsplot}, \ref{Fepsfit}). The wave dynamics of the examined structure is studied through the numerical procedure described in Sect. \ref{model}.
Fig.  \ref{waveplot1d} shows the propagation of compression solitary waves under the application of an initial velocity $v_0=5.0$ m/s to the left base. The traveling pulses exhibit compact support spanning 6-7 units, in agreement with the theory presented in Sect. \ref{stiffening}.
Previous results on the compression solitary wave dynamics of larger scale tensegrity chains can be found in \cite{25}.
 
\section{Two-dimensional beams}
\label{sec:2Dwaves}

The present section illustrates impact simulations on different assembles of the 2D square cells illustrated in Fig. \ref{assemblages}-Top, making use of the geometric and material properties reported in Table~\ref{tab:1}.  
The analyzed systems consist of a  series of beams (or strips), which are formed by 3$\times$50 and 1,3,5$\times$80 grids of square cells. We will see that the wave dynamics of such systems retains the distinctive feature of the response observed in the 1D chain of Sect. \ref{sec:1Dwaves}, due to the formation and propagation of compact compression waves under impact loading. Nevertheless, the response of the 2D systems examined hereafter is accompanied by dynamical events not observable in 1D mass-spring systems, which originate from the onset of a diffuse agitation motion of the lattice around the reference configuration, near the load application points, reflection edges and collision regions (`thermalization effect' \cite{mams2010}). 

\subsection{Effects of different impact velocities}

\begin{figure*}[htbp]
\begin{center}
\includegraphics[width=\linewidth,angle=0]{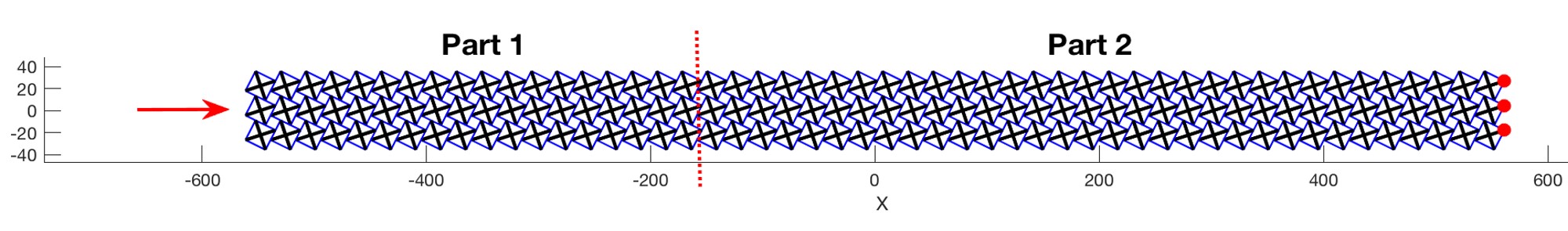}
\caption{A $3\times50$ strip of square cells, with the nodes of the right base fixed and the four nodes of the central cel facing the left edge subject to an imposed initial velocity $v_0$ along the longitudinal direction.
Part 1 is composed of $3\times18$ cells, while Part 2 is composed of $3\times32$ cells.}\label{strip1}
\end{center}
\end{figure*}

Let us begin with the analysis of the 3$\times$50 strip illustrated in Fig.~\ref{strip1}. We divide such a system into a first section of $3\times18$ cells, and a second section of $3\times32$ cells for post-processing purposes, as explained hereafter. The nodes of the right base are supposed to be at rest, while the four nodes of the central unit facing the left base are assigned a prescribed horizontal initial velocity with amplitude $v_0$. The following results examine the wave dynamics of the system shown in Fig.~\ref{strip1} for different values of $v_0$, which allow us to study the propagation and reflection of the traveling pulses within few tenths of a second.

\begin{figure*}[htbp]
\begin{center}
\includegraphics[width=1.0\linewidth,angle=0]{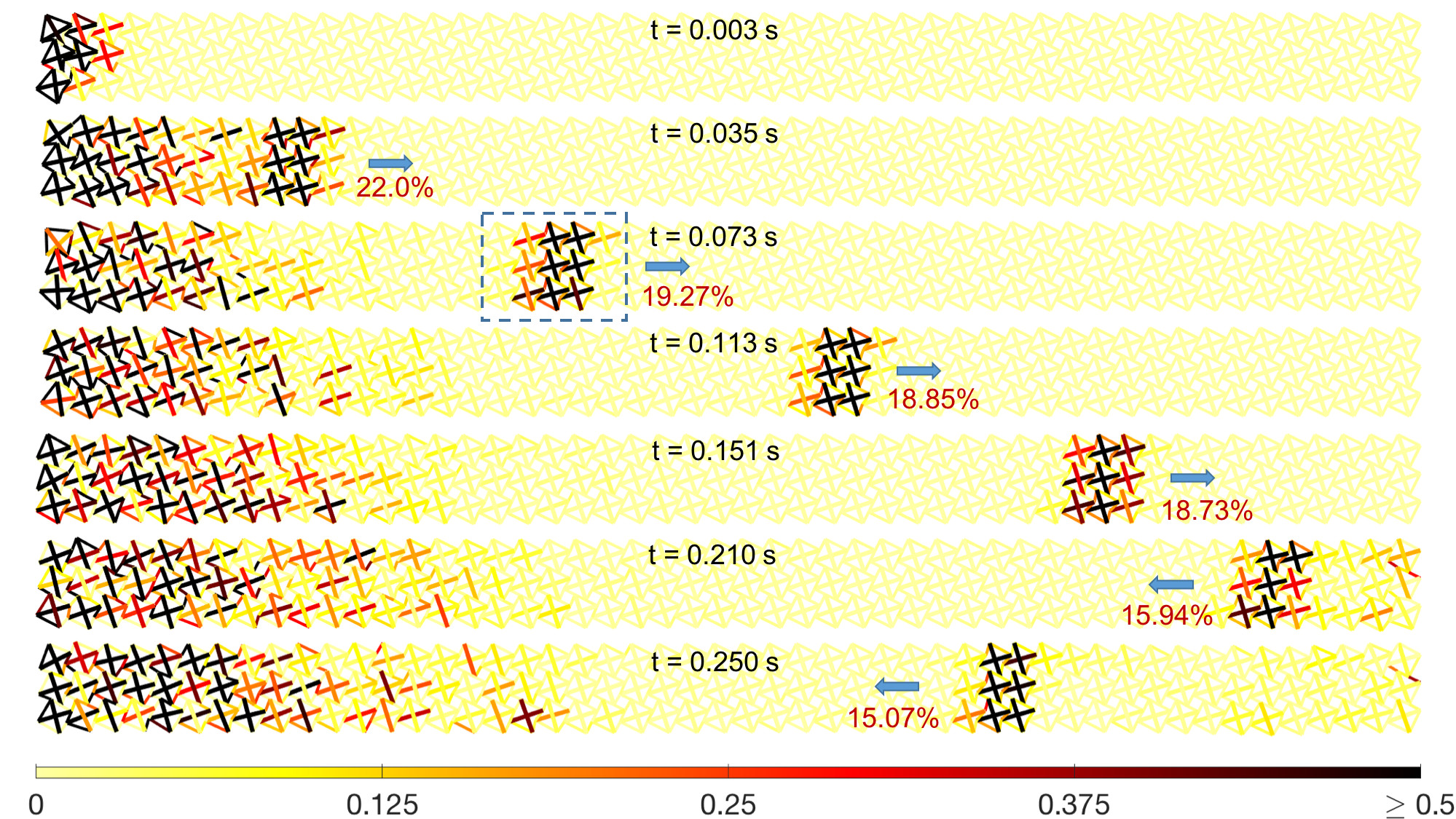}
    \caption{Deformed configurations with superimposed colormaps of the energy carried by the different members for a 3$\times$50 strip at different times after impact with initial velocity $v_0=5$\,m/s. Energy values are expressed as energy fractions (\%) of the total energy carried by the strip.
    The red numbers in front of the ccw indicate the total energy fraction $\hat{E}_{ccw}$ transported by the localized wave.}\label{strip2a}
\end{center}
\end{figure*}

With reference to the impact simulation with $v_0=5$\,m/s, we illustrate in Fig.~\ref{strip2a}  the deformed configurations of the structure with superimposed colormaps of the total energy carried by the different members at different times, before and after the reflection of the traveling pulses at the fixed end. In such a figure and the remainder of the paper, the total energy carried by a bar element is computed as the summation of its elastic energy, $\frac{1}{2}\,\tilde k_i\,(l_i(\pb) - \bar l )^2$, plus the kinetic energy of the two  nodes attached to the bar. The energy carried by the cable elements is instead identified with the competent elastic energy, due to the very light mass of these members. By dividing the members' energies  by the total energy carried by the strip, we are led to the \textit{energy fractions} graphically illustrated by the colormaps depicted in Fig.~\ref{strip2a}.
The results presented in such a figure and the animations given in Appendix highlight two main mechanisms of propagation of the applied compressive disturbance: $i)$ `thermalization' of the lattice \cite{mams2010}), due to the onset of an agitation motion in the region of the strip located behind the point of application of the impact load (Part 1); $ii)$ separation and propagation of a compact compression wave (ccw), whose support  consists of a packet of 6/9 units (about three columns of cells), in front of the thermalized region (Part 2).

The snapshots given in Fig.~\ref{strip2a}, as well as those included in the subsequent figures, quote the total energy fraction $\hat{E}_{ccw}$ transported by the ccw at different times (red numbers in front of the ccw). Such a quantity is defined as the summation of the energy fractions competing to the lattice members that form five lattice spaces embracing the current configuration of the ccw (cf. the dashed box in the panel for $t= 0.073$ s).  The given definition of $\hat{E}_{ccw}$ allows us to account for second-order effects, since one observes that most of the ccw energy  is localized in three lattice spacings, as  we already noticed. The quantity $\hat{E}_{ccw}$ is proportional to the square of the amplitude of the ccw, and we see from  Fig.~\ref{strip2a} that this quantity undergoes a 2.8 \% decrease while traveling from the impacted base to the fixed edge (see the relative difference between the values of $\hat{E}_{ccw}$ at $t=0.151$ s and $t=0.073$ s), and $\approx 15$ \% decrease after reflection of the ccw at the fixed base (relative difference between the values of $\hat{E}_{ccw}$ at $t=0.210$ s and $t=0.151$ s). 
The small decrease of the ccw energy before reflection is explained by some light leaking effects of the traveling wave. The more significant reduction of such a quantity after reflection of the ccw is instead explained by a thermalization phenomenon generated by the wave reflection near the fixed base. Clearly the energy lost by the ccw gets smeared over the lattice, due to the conservative nature of the examined beam.

In order to better understand the wave dynamics of the system under consideration, we applied different impact velocities to the strip depicted in Fig.~\ref{strip1}, and numerically computed the velocity $V_{ccw}$ exhibited by the center of mass of the ccw during the interval intercurring between the time $t_a$ at which the ccw has fully entered Part 2, and the time $t_b$ at which the right end of the ccw has reached the fixed base. We characterized the variation of $V_{ccw}$ in such a time window through numerical predictions of the mean value $\bar{V}_{ccw}$ and the standard deviation $s_{{V}_{ccw}}$ of such a quantity.
We also computed the maximum strain $\varepsilon_m$ suffered by the cables, and the relative variation $\delta \hat{E}_{ccw} = (\hat{E}_{ccw}^{(a)}-\hat{E}_{ccw}^{(b)})/\hat{E}_{ccw}^{(a)}$ of the ccw energy in the  $[t_a, t_b]$ time window.
The results shown in Tab. \ref{tab:2} reveal that the velocity and the energy of the ccw remain nearly constant during the propagation of such a wave across Part 2 of the system, when $v_0$ varies from 2.50 m/s up to 5.00 m/s. 
The variations of $V_{ccw}$  and $\hat{E}_{ccw}$ in the time interval $[t_a, t_b]$ are  
rather small, being less than 1 \% in terms of velocity, and less than 2 \% in terms of energy. 
It is worth noting that the maximum cable strain $\varepsilon_m$ approximatively reaches the yield strain of the employed Nylon  12 fibers (cf. Sect. \ref{model}) for $v_0= 5.00$ m/s.
We also observe from Table~\ref{tab:2} that the ccw phase speed depends in a nonlinear fashion on the maximum strain $\varepsilon_m$ suffered by the strings, which is qualitatively in agreement with the predictions
of the theory of nonlinear waves recalled in Sect. \ref{stiffening} (see Eqn. \eqref{eq:Vs}).

\begin{table}
\caption{Statistics of the wave speed, maximum axial strain of the cables, and total energy fraction of the ccw traveling across Part 2, for different impact velocities.}
\label{tab:2}       
\begin{tabular}{ccccc}
\hline\noalign{\smallskip}
$v_0$ & $\bar{V}_{ccw}$  & $s_{{V}_{ccw}}$  & $\varepsilon_m$ & $\Delta \hat{E}_{ccw}$ \\
 (m/s) &  (m/s) & (m/s)  &  (\%) &  (\%) \\
\noalign{\smallskip}\hline\noalign{\smallskip}
5.00 & 5.67 & 0.02 & 33.46 & 1.92\\ 
3.75 & 5.10 & 0.02 &  18.41 & 0.87\\ 
2.50 & 4.27 & 0.02 &  8.09 & 0.65\\ 
\noalign{\smallskip}\hline
\end{tabular}
\end{table}

\begin{figure}[htbp]
\begin{center}
\includegraphics[width=0.99\linewidth,angle=0]{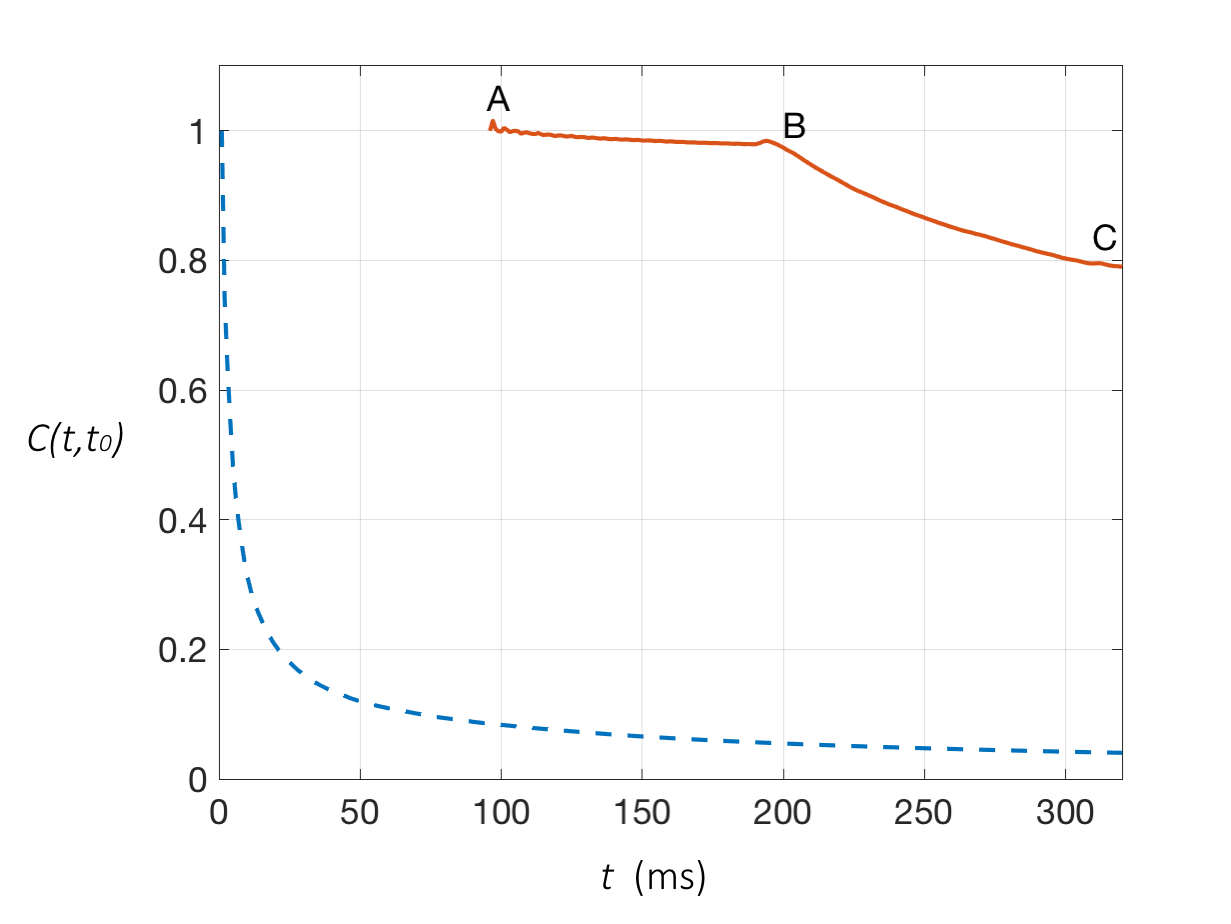}
    \caption{Energy correlation function in Part 1 (dashed blue curve) and Part 2 (solid red curve) for $v_0=5$\,m/s.}\label{strip2encorr}
\end{center}
\end{figure}

We further analyze the partition of the impact energy on examining the time-variation of the energies $E_j(t)$ that are associated with the nodes of Part 1 and Part 2. The latter are computed by adding the kinetic energy competing to the generic node $j$ to one half of the summation of the elastic energies carried by all the members $i$ that are attached to node $j$, obtaining
\begin{equation}
E_j(t)=\frac{1}{2}m_j v_j^2(t)+\frac{1}{4}\sum_{i} \ \tilde k_i (l_i(t)-\bar l_i)^2
\label{enode}
\end{equation}
The study of the temporal evolution of the nodal energies is conducted by introducing the following energy correlation function \cite{mams2010}
\begin{equation}
C(t,t_0)=\frac{c(t)}{c(t_0)},
\label{corr0}
\end{equation}
with
\begin{equation}
c(t)=\frac{1}{N_{b}\!-\!N_{a}}\left\langle \sum_{i=N_a}^{N_b} E_i^2(t)\!\!\right\rangle \,-\, \left\langle\!\!\frac{1}{N_a\!-\!N_b} \sum_{i=N_a}^{N_b} E_i(t)\!\!\right\rangle^{\!\!\!2}
\label{corr1}
\end{equation}
Here, the angled brackets denote temporal averages from time $t_0 $ to current time $t$ \cite{mams2010}, while $N_a$, $N_b$ denote the first and last nodes of the part under consideration. For Part 1, we take $t_0=0$, while for Part 2 we take $t_0=t_a$.
Figure~\ref{strip2encorr} shows the energy correlation function in Part 1 (dashed blue curve), and Part 2  (solid red curve) for $v_0 = 5$ m/s. Point A corresponds to $t=t_a$, while point B corresponds to $t=t_b$ (final time before reflection of the ccw at the fixed edge). Finally, point C corresponds to the time at which the reflected ccw exits from Part 2. The results in Fig.~\ref{strip2encorr} highlight marked energy equipartition in Part 1 (correlation function tending to zero), and, conversely, marked energy localization in Part 2 (correlation function close to one), up to the reflection of the ccw at the fixed edge. Some light themalization of Part 2 is also observed after reflection of the ccw at the fixed edge, which corresponds to the decreasing branch B-C of the solid red curve  in Figure~\ref{strip2encorr} (see also the two bottomost panels of Fig. \ref{strip2a}).

\subsection{Collision between two compact compression waves}

\begin{figure*}[htbp]
\begin{center}
\includegraphics[width=1.0\linewidth,angle=0]{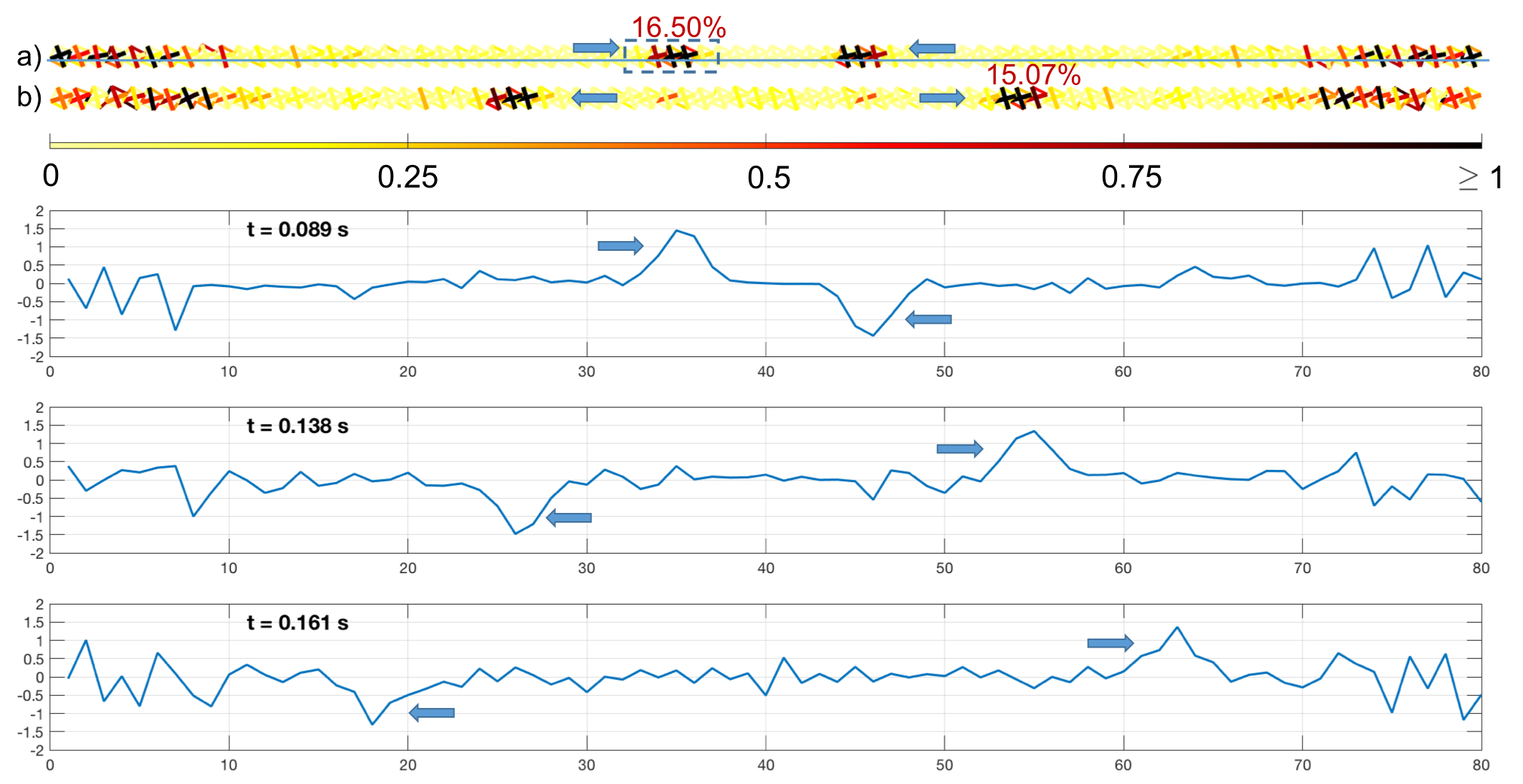}
\caption{Top:
Deformed configurations with superimposed colormaps of the members' energy fractions for a 1$\times$80 strip at different times after double impact with initial velocity $v_0=5$\,m/s: a) t=0.089 s; b) t=0.138 s.
Bottom: 
Plots of the horizontal velocities exhibited by the nodes on the longitudinal line shown in the top panel (a) (positive values for velocities directed from left to right).}\label{en_profile1x80v5}
\end{center}
\end{figure*}

\begin{figure*}[htbp]
\begin{center}
\includegraphics[width=1.0\linewidth,angle=0]{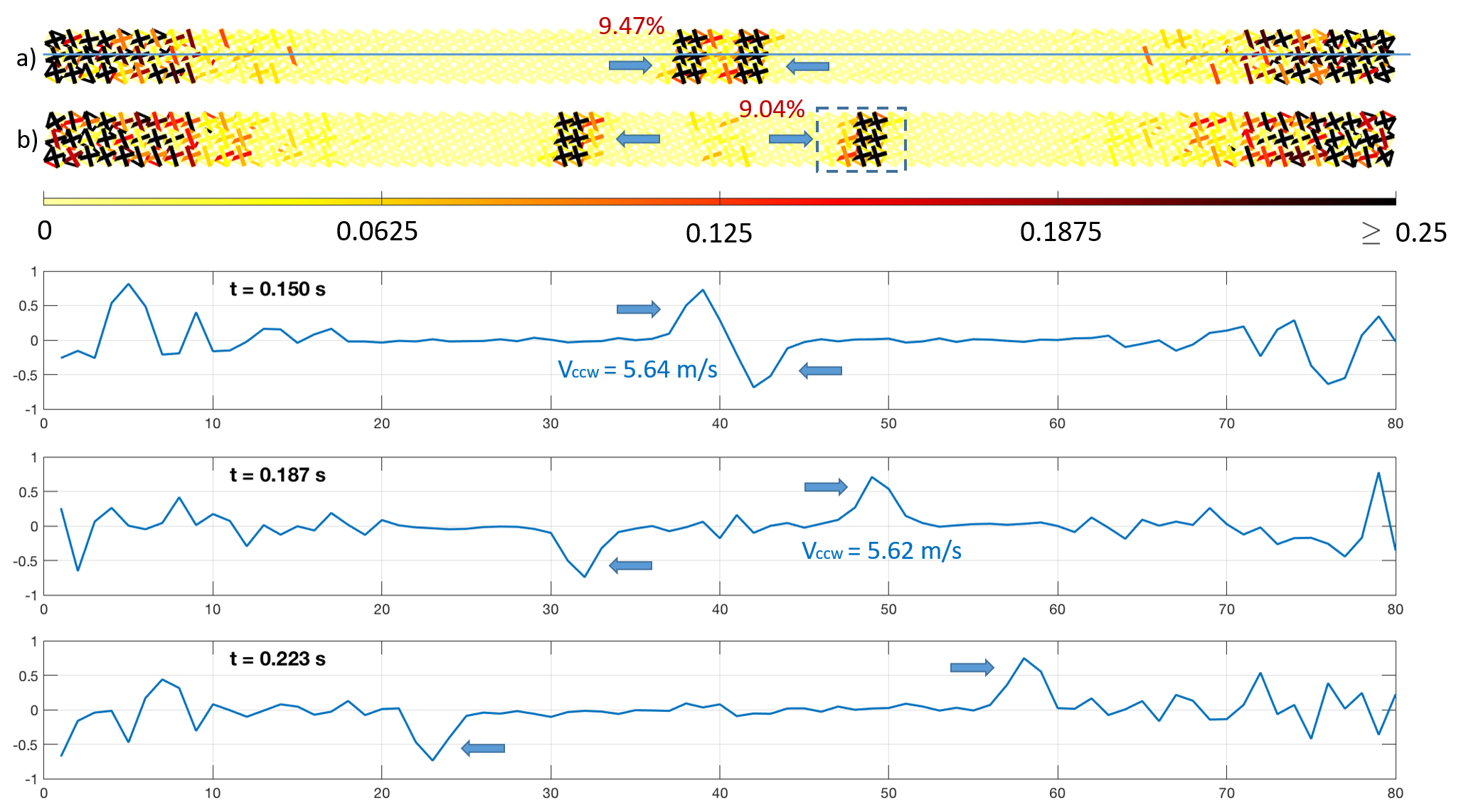}
\caption{Top:
Deformed configurations with superimposed colormaps of the members' energy fractions for a 3$\times$80 strip at different times after double impact with initial velocity $v_0=5$\,m/s: a) t=0.150 s; b) t=0.187 s.
Bottom: 
Plots of the horizontal velocities exhibited by the nodes on the longitudinal line shown in the top panel (a) (positive values for velocities directed from left to right).}
\label{en_profile3x80v5}
\end{center}
\end{figure*}

\begin{figure*}[htbp]
\begin{center}
\includegraphics[width=1.0\linewidth,angle=0]{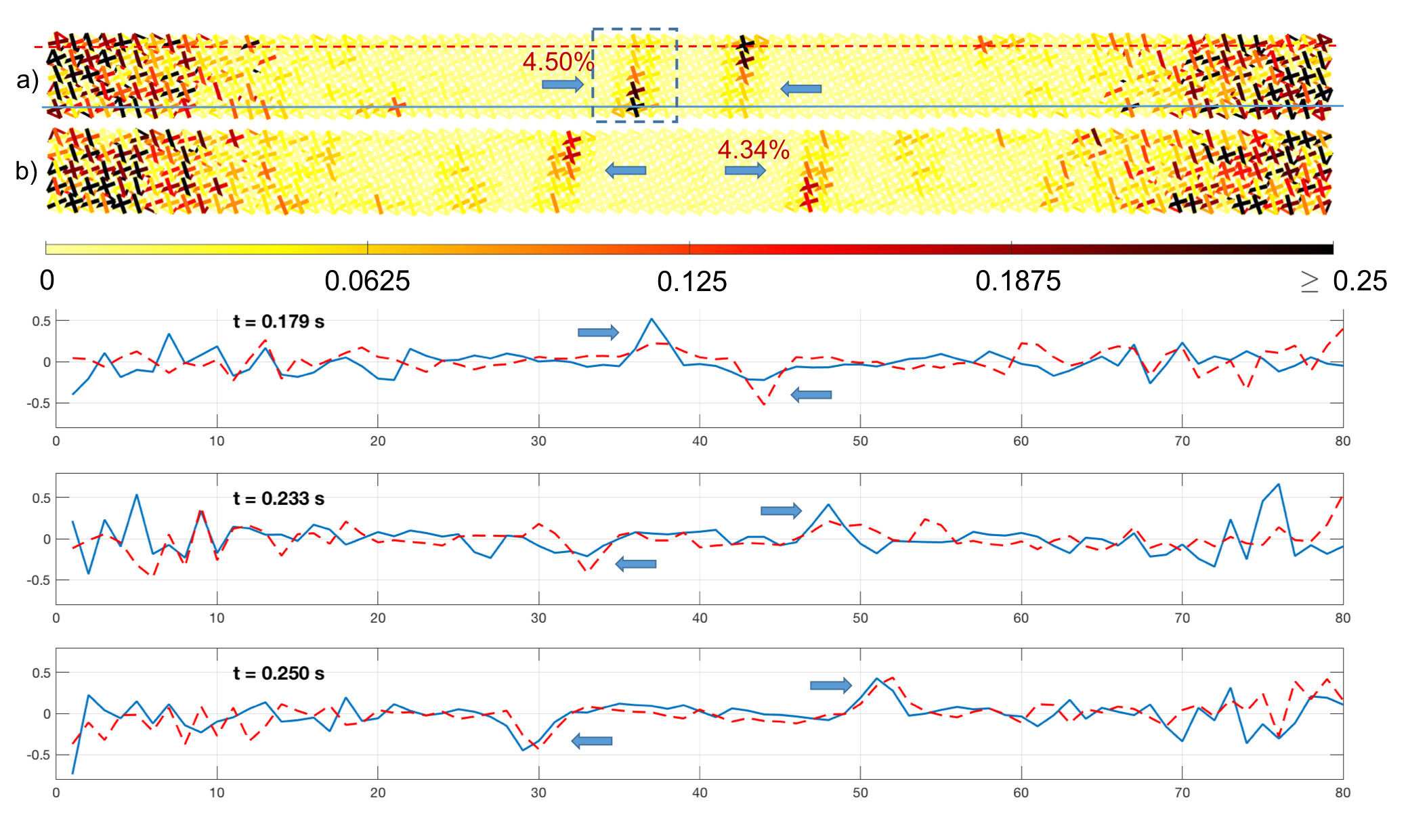}
\caption{Top:
Deformed configurations with superimposed colormaps of the members' energy fractions for a 5$\times$80 strip at different times after double impact with initial velocity $v_0=5$\,m/s: a) t=0.179 s; b) t=0.233 s.
Bottom: 
Plots of the horizontal velocities exhibited by the nodes of the bottom sideline (solid blue curve) and top sideline (dashed red curve, positive values for velocities directed from left to right). The examined nodes are located on the longitudinal lines shown in the top panel (a).}
\label{en_profile5x80v5}
\end{center}
\end{figure*}

It is known that compactons emerge unmodified after collision with other compactons, with exception to a phase shift (refer, e.g., to \cite{Rosenau2005}). In order to numerically investigate on the `compacton' nature of the ccws traveling in 2D beams of square cells, 
we studied the wave dynamics of strips formed by 1$\times$80, 3$\times$80, and 5$\times$80 cells, which are symmetrically impacted on both ends with initial velocity $v_0=5$ m/s. The examined systems have the terminal bases free from constraints and are such that the four nodes of the central units facing  such bases are assigned initial velocities equal to $v_0$.
Figs. \ref{en_profile1x80v5}-\ref{en_profile5x80v5} show
the deformed configurations with superimposed colormaps of the members' energy fractions for the systems under examination,
together with the values of the total energy fraction $\hat{E}_{ccw}$ transported by the ccw (red numbers in front of the ccws). Different configurations are analyzed, before and after the collision of the ccws emerging from the thermalized regions near the impacted ends. Plots of the horizontal velocities exhibited by the nodes of selected longitudinal segments are included in the bottom panels of the above figures (nodal velocity profiles).
The results presented in Figs. \ref{en_profile1x80v5}-\ref{en_profile5x80v5} highlight a decrease of $\hat{E}_{ccw}$ after collision, which reduces in amplitude with the thickness of the system, beaing respectively equal to 8.7 \%, 4.5 \% and 3.6 \% in the 1$\times$80, 3$\times$80 and 5$\times$80 beams.  
The nodal velocity profiles shown in the bottom panels of such figures show positive and negative peaks exhibiting nearly constant values over time. 
In particular, with reference to the 3$\times$80 strip we notice that the velocity ${V}_{ccw}$ of the ccw traveling from left to right exhibits a very small variation in the interval delimited by a time preceding the collision (${V}_{ccw}=5.64$ m/s at $t=0.150$ s), and a time following the collision (${V}_{ccw}=5.62$ m/s at $t=0.187$ s, cf. Fig. \ref{en_profile3x80v5}). An analogous result characterizes the propagation of the ccw traveling in the opposite direction.
We also observe an asymmetric distribution of the traveling energy across the thickness of the ccw (look at the different colors of the members forming the cells at the wavefront, especially in the 3$\times$80, and 5$\times$80 systems), which is due to the chiral aspect of the examined lattice.
This asymmetry of the wavefront produced peaks of the transported energy close to the lateral edges of the strip, far from  the centerline (wave `drifting' effects).
Some light or moderate thermalization effects can be observed in the central portion of the strip, after the collision of the two ccws (observe the colormaps of the members' energy fractions, and the longitudinal wave profiles at times greater than the collision time in Figs. \ref{en_profile1x80v5}-\ref{en_profile5x80v5}).  Such a central thermalization of the lattice decreases in magnitude by increasing the thickness of the strip (that is, when passing from the 1$\times$80 to the 3$\times$80 and 5$\times$80 systems), and produces small amplitude oscillations of the longitudinal velocity profiles.
In the the 3$\times$80 and 5$\times$80 systems, one notes a decrease of the amplitude of such velocity oscillations with time (cf. the bottom panels of Figs. \ref{en_profile3x80v5}-\ref{en_profile5x80v5}). 

We close the present section by remarking that the results shown in 
Fig. \ref{strip2a} and Figs. \ref{en_profile1x80v5}-\ref{en_profile5x80v5} highlight some rearrangements of the energy distribution at the wavefront of the ccws (look at the energy fractions carried by the different members forming the ccw, at different times). This implies that the spatial shape of such waves is not fully conserved during their propagation across the examined systems. Nevertheless, in a given beam and under a prescribed impact velocity, the support-size and the speed of the traveling ccws appear to remain substantially unaltered, during the propagation and collision of such waves, up to small relative errors. 
It is also worth remarking that the thickness of the beam influences both the thermalization effects induced by the collision of two ccws (as we already observed), and the the size of the support of the ccws.  The latter indeed appear to shrink in the longitudinal direction, and to increase in the transverse direction, when passing from the 1$\times$80 strip (Fig. \ref{en_profile1x80v5}) to the 5$\times$80 strip (Fig. \ref{en_profile5x80v5}).

\section{Three-dimensional beams} \label{sec:3Dwaves}

\begin{figure*}[htbp]
\begin{center}
\includegraphics[width=\linewidth,angle=0]{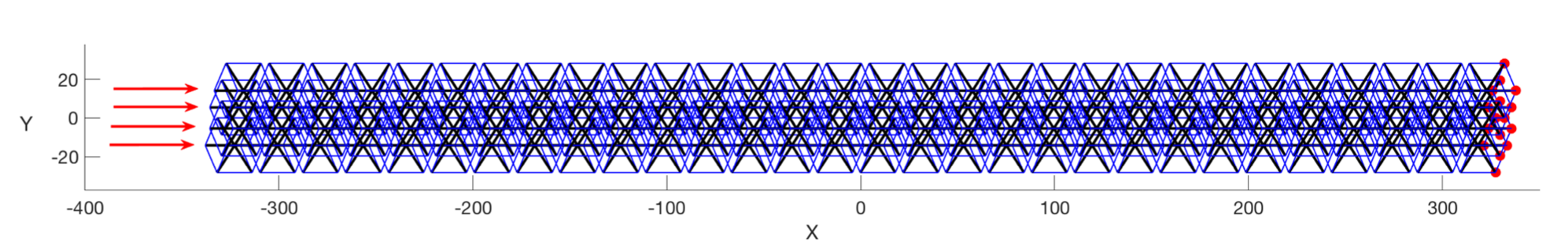}
\caption{A $2\times2\times30$ assembly of cubic cells, which features fixed nodes at the right base (red dots) and is impacted through application of an initial velocity $v_0=1.25$\,m/s to the nodes of the four cells facing the left base (red arrows).}\label{setup3d}
\end{center}
\end{figure*}

\begin{figure*}[htbp]
\begin{center}
\includegraphics[width=0.92\linewidth,angle=0]{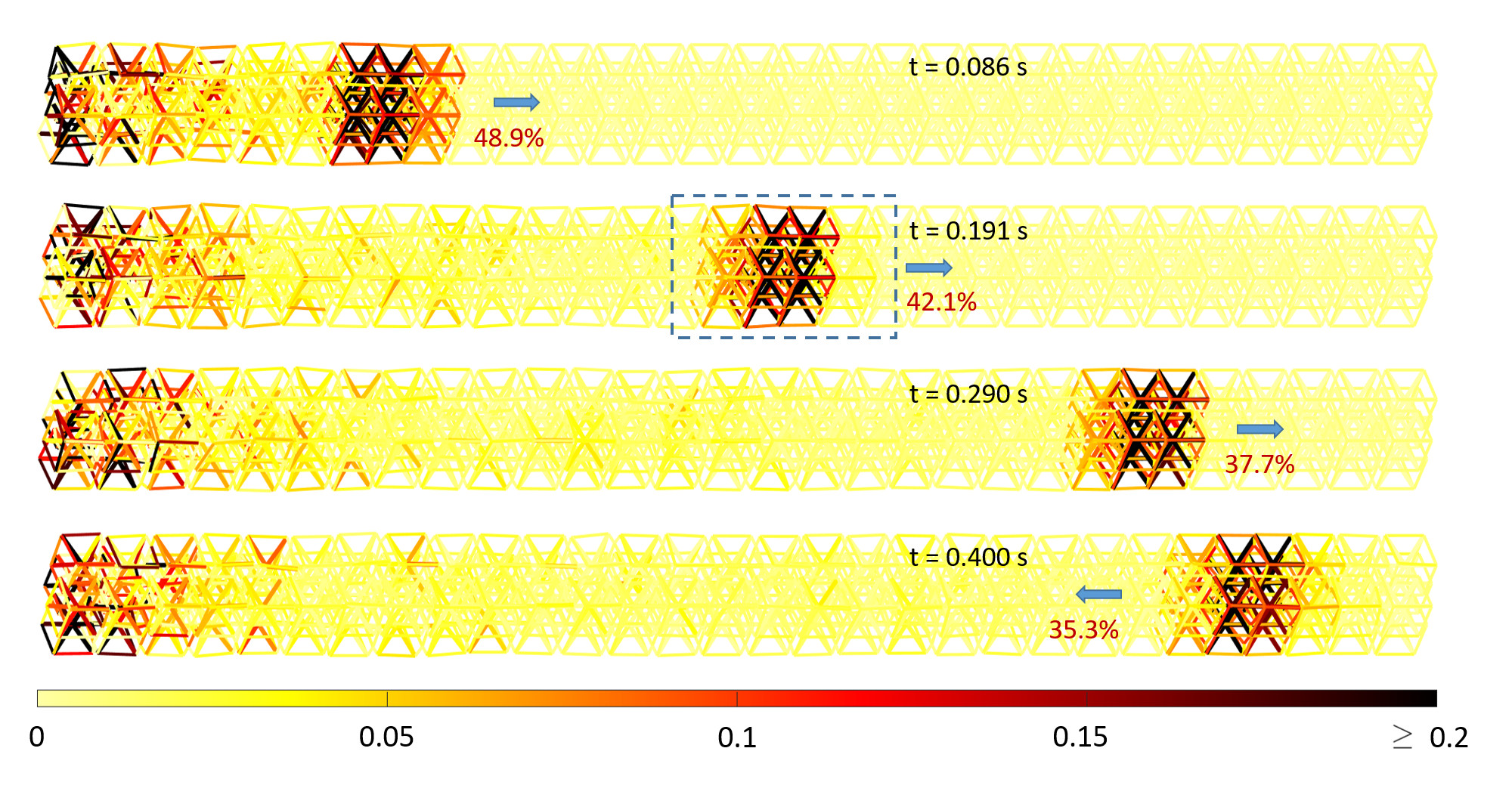}
\caption{
Deformed configurations with superimposed colormaps of elements' energy fractions at different times after impact with initial velocity $v_0=1.25$\,m/s, on a $2\times2\times30$ beam of cubic cells.}
\label{energy3d}
\end{center}
\end{figure*}

The 3D systems examined in the present study consist of $2\times2\times30$ beams formed by cubic cells, which are subject to different impact loading conditions  (cf. Figs.~\ref{assemblages}-Bottom and Fig.~\ref{compression3d}). Let us first study the wave dynamics of the beam shown in Fig.~\ref{setup3d}, under the application of an initial velocity $v_0=1.25$ m/s to the nodes of the four cells facing the left base. 
Fig. \ref{energy3d} shows the deformed configurations of such a beam with superimposed colormaps of members' energy fractions, at different times after impact. As in the 2D lattices studied in Sect. \ref{sec:2Dwaves}, also in the current system we notice the formation of a ccw separating from a thermalized zone facing the impacted base. 
However, in the 3D case of Fig. \ref{energy3d}, we observe that such a thermalized region exhibits slightly smaller longitudinal extension, as compared to the 2D beam studied in Fig. \ref{strip2a}. One notices that the total energy fraction transported by the 3D ccw is $\approx 38$ \% (before the impact with the fixed end), while it results $\hat{E}_{ccw}\approx 19$ \% in the 2D example shown in Fig. \ref{strip2a}.
Opposite trend shows the temporal variation of $\hat{E}_{ccw}$,
which is larger in the current 3D beam than in the 2D beam of Fig. \ref{strip2a}. Referring to the 3D case in Fig. \ref{energy3d}, we note that the decrease of $\hat{E}_{ccw}$ amounts to 22.90 \% over the time window  [0.086, 0.290] s, and to 10.45 \% over the interval [0.191, 0.290] s. The decrease of total energy fraction due to the reflection of the 3D ccw at the fixed base is equal to 6.37 \%, over the time window [0.290, 0.400] s.

\begin{figure*}[htbp]
\begin{center}
\includegraphics[width=0.88\linewidth,angle=0]{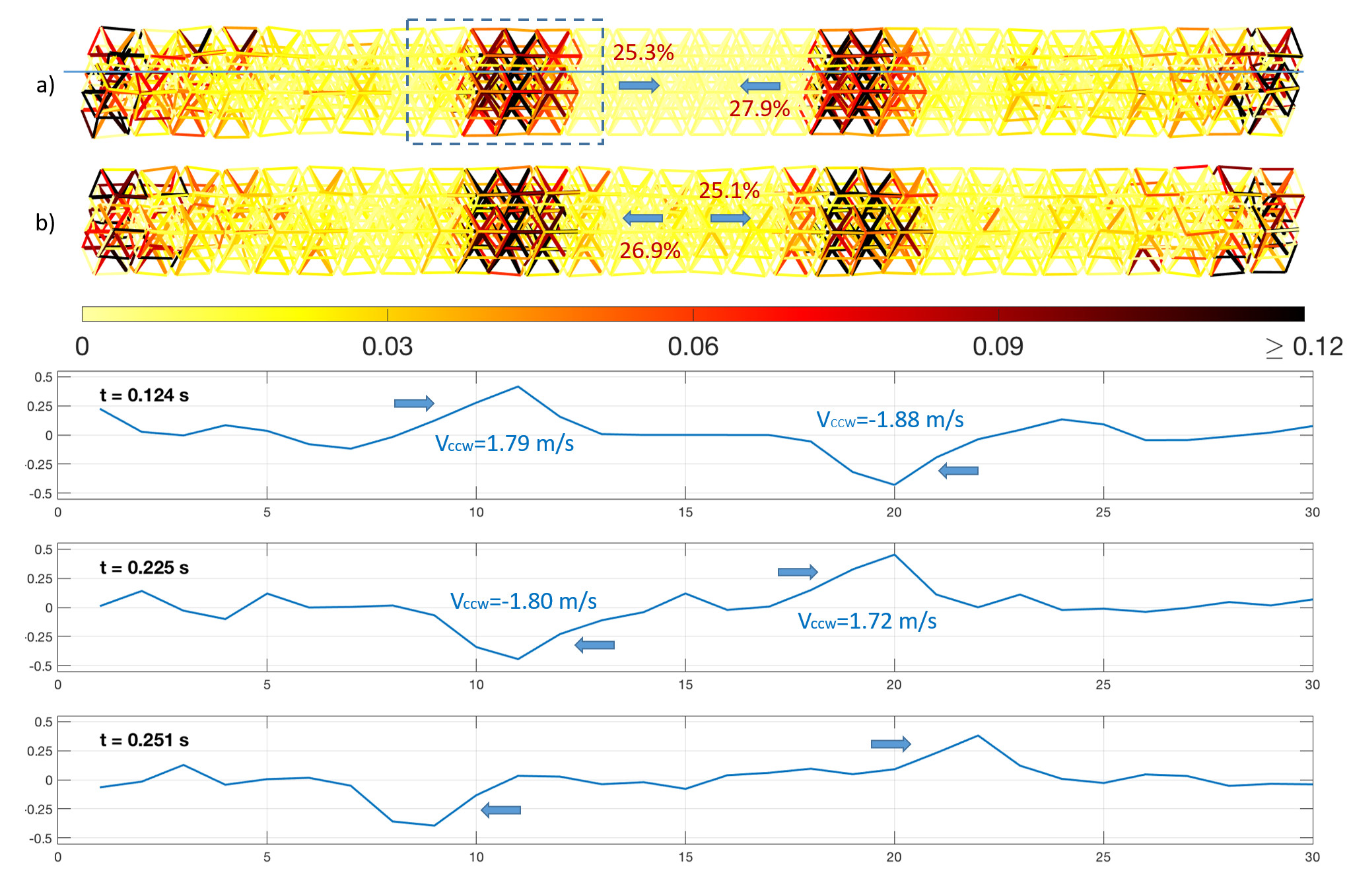}
\caption{
Top:
Deformed configurations with superimposed colormaps of members' energy fractions for a $2\times2\times30$ beam of cubic cells at different times after double impact with initial velocity $v_0=1.25$\,m/s: a) t=0.124 s; b) t=0.225 s.
Bottom: 
Plots of the horizontal velocities exhibited by the nodes on the longitudinal line shown in the top panel (a).}
\label{en_profile3d}
\end{center}
\end{figure*}

The collision of two 3D ccws is analyzed in Fig. \ref{en_profile3d}, which illustrates a double impact simulation on a $2\times2\times30$ beam of cubic cells ($v_0=1.25$\,m/s). The results in Fig. \ref{en_profile3d} highlight the formation of two  ccws near the impacted bases, which travel in opposite directions with approximatively constant velocity, before and after collision. Some light thermalization effects are observed in the central portion of the beam after the collision of the two ccws, which generate oscillatory pulses behind the leading waves in the longitudinal velocity profiles. This thermalization phenomenon also produces a slight decrease of $\hat{E}_{ccw}$ before and after the collision of the ccws. One however observes that the peaks of the velocity profiles comprised between the leading pulses  progressively decrease in amplitude with time (cf. Fig. \ref{en_profile3d}-Bottom). It is worth emphasizing that the two opposite ccws are not perfectly symmetric in the 3D beam, due to the fact that the two halves of such a system separated by the vertical centerline are not mirror copies of one another (cf. Fig.~\ref{setup3d}).

\section{Concluding remarks} \label{conclusions}

We have numerically studied the compact wave dynamics of multidimensional tensegrity beams with nonlinear behavior induced by a stiffening-type elastic response. An impact simulation conducted on a 1D mass-spring system has allowed us to extend previous results on the compression solitary wave dynamics of tensegrity chains \cite{25} at a lower-scale. A parade of numerical simulations of impact events on 2D and 3D tensegrity beams has led us to discover, for the first time, that the impact dynamics of such systems is characterized by the combination of thermalization phenomena in proximity of the impacted areas, and the formation and propagation of compact compression waves in front of the thermalized regions. The traveling ccws transport energy on localized packets of unit cells spanning from two to three lattice modules in the longitudinal direction. Such compression waves propagate with nearly constant velocity before and after collisions with other ccws, and exhibit limited energy leaking during their propagation. The observed thermalization effects are marked in proximity of the regions of application of the impact loads, and lighter in correspondence with the collision zones of ccws traveling in opposite directions. Some rearrangements of the members' energy fractions are observed at the wavefront during the propagation of such waves. Overall, we are led to conclude that the compact compression waves carried by the examined tensegrity beams approximatively retain the  size of their support and the ccw phase speed, while exhibiting light modifications of the spatial shape and the transported energy, during propagation and collisions with other ccws (`quasi-compactons').

The behaviors examined in the present study suggest the employment of tensegrity beams to form innovative acoustic lenses. Such devices are expected to be able to generate tunable ccws in an adjacent host medium, which will cohalesce at a given focal point \cite{spadoni}. The latter may consist of  a material defect or a tumor mass in the host medium \cite{tensegritypatent}.  
Compared to devices based on granular metamaterials supporting fixed wavelength solitary waves \cite{spadoni}, the tensegrity acoustic lenses will profit from the  adjustable width of ccws in tensegrity lattices (refer, e.g., to the observations made at the end of Sect. \ref{sec:3Dwaves}), and the atomic-scale localization phenomenon observed in the high-energy limit \cite{25}. We address such a pioneering approach to sound focusing to future work. Additional future research lines 
will include the analytic modeling of the wave dynamics of tensegrity lattices, through multiscale approaches aimed at developing a general quasicontinuum nonlinear partial differential equation of periodic systems \cite{Nesterenko_2001,Rosenau1987}. 
This study will allow us to further investigate on the compacton nature of the mechanical waves supported by tensegrity lattices in the continuum limit.
We also intend to experiment novel additive manufacturing techniques for the fabrication of macro- and micro-scale physical models of lattices with tensegrity architecture, through future work. Such mockups will be subject to dynamical tests aimed at experimental validating the puzzling compact wave dynamics of highly nonlinear tensegrity metamaterials, at multiple scales.

\section*{Acknowledgements}
This is a pre-print of an article published in Nonlinear Dynamics, Springer (ISSN: 0924-090X). The final authenticated version is available online at: https://doi.org/10.1007/s11071-019-04986-8 .

AM and GR gratefully acknowledge the financial support from the Italian Ministry of Education, University, and Research (MIUR) under the `FFABR' grant L.232/2016. FF gratefully acknowledges financial support from the Italian Ministry of Education, University, and Research (MIUR) under the `Departments of Excellence' grant L.232/2016.

%



\end{document}